# Automatic Multi-Objective Optimization of Coarse-Grained Lipid Force Fields Using *SwarmCG*

Charly Empereur-mot,[a,*] Riccardo Capelli,[b] Mattia Perrone,[b] Cristina Caruso,[b] Giovanni Doni[a] & Giovanni M. Pavan [a,b,*]

[a] *Department of Innovative Technologies, University of Applied Sciences and Arts of Southern Switzerland, Polo Universitario Lugano, Campus Est, Via la Santa 1, 6962 Lugano-Viganello, Switzerland*

charly.empereur-mot@supsi.ch

[b] *Politecnico di Torino, Department of Applied Science and Technology, Corso Duca degli Abruzzi 24, Torino, 10129, Torino, Italy*

giovanni.pavan@polito.it

**ABSTRACT** The development of coarse-grained (CG) molecular models typically requires a time-consuming iterative tuning of parameters in order to have the approximated CG models behaving correctly and consistently with, *e.g.*, available higher-resolution simulation data and/or experimental observables. Automatic data-driven approaches are increasingly used to develop accurate models for molecular dynamics simulations. But the parameters obtained *via* such automatic methods often make use of specifically-designed interaction potentials, and are typically poorly transferable to molecular systems or conditions other than those used for training them. Using a multi-objective approach in combination with an automatic optimization engine (*SwarmCG*), here we show that it is possible to optimize CG models that are also transferable, obtaining optimized CG force fields (FFs). As a proof of concept, here we use lipids, for which we can avail of reference experimental data (area per lipid, bilayer thickness) and reliable atomistic simulations to guide the optimization. Once the resolution of the CG models (mapping) is set as an input, *SwarmCG* optimizes the parameters of the CG lipid models iteratively and simultaneously against higher-resolution simulations (*bottom-up*) and experimental data (*top-down* references). Including different types of lipid bilayers in the training set in a parallel optimization guarantees the transferability of the optimized lipid FF parameters. We demonstrate that *SwarmCG* can reach satisfactory agreement with experimental data for different resolution CG FFs. We also obtain stimulating insights on the precision-resolution balance of the FFs. The approach is general and can be effectively used to develop new FFs, as well as to improve existing ones.

## I. INTRODUCTION

Molecular dynamics (MD) has become a fundamental tool in the study of complex molecular systems, providing high-resolution insights often inaccessible via experimental techniques. One of the main limitations of all-atom (AA) MD simulations is the space and time scales accessible with current computational capabilities. Coarse-grained (CG) molecular models, in which groups of atoms are represented as larger CG particles (or beads), may alleviate these issues and are increasingly employed to study systems of interest in structural biology[1–3], drug discovery[4,5], biophysics[6–8] and nanomaterials design[9–11]. Lipid bilayers, key components of the cell membranes, are a notable example of supramolecular systems exhibiting properties which, to a large extent, require CG models to be studied effectively[12].

Different approaches are typically used for the development of CG force fields (FFs)[13–26]. *Top-down* strategies essentially aim at reproducing molecular properties observed experimentally with the CG models. Conversely, *bottom-up* approaches rely on calibrating CG model parameters using equilibrium simulations of higher-resolution molecular models (*e.g.*, AA). Widely used for the simulation of lipids, the popular Martini[24] FF, for example, presents aspects of both. The possibility to parametrize molecular models for a variety of molecules using transferable CG beads makes the Martini[24] FF versatile. However, general CG FFs remain intrinsically approximated in modeling specific molecular systems compared to CG models that, albeit less general and transferable, are optimized *ad hoc* to this end[18,27].

Methods such as, *e.g.*, Inverse Monte Carlo (IMC)[13], Iterative Boltzmann Inversion (IBI)[15], Multi-State IBI (MS-IBI)[19], Force Matching (FM)[16–18,28], ForceBalance[29,30], Relative Entropy Minimization (REM)[31], the generalized Yvon-Born-Green (g-YBG)[32] equation, and different flavors of Particle Swarm Optimization (PSO)[21,22,33,34] have been used as basis to build *bottom-up* and/or *top-down* FF parametrization approaches to calibrate AA FFs relying, *e.g.*, on quantum mechanical data[30,35,36], or to calibrate CG FFs based on AA MD trajectories[37–42,34,43,44]. For what pertains to the *bottom-up* route, their parameters extraction schemes are based on, *e.g.*, reproducing pair distribution functions[13,15,19], matching forces[16–18,28–30], minimizing the information loss in terms of relative entropy[31], or on the liquid state theory[32] (see, *e.g.*, Kmiecik *et al.*[45] for an exhaustive review of these approaches).

More recently, the evolution of machine learning approaches is considerably accelerating the development of accurate CG molecular models. Deshmukh *et al.* developed a CG FF for different solvents[21,22], hydrocarbons[46], small peptides[33] and several polymers[47], optimizing the interaction parameters to reproduce experimental observables (exclusively *top-down*) using particle swarm optimization[48] (PSO) and artificial neural networks (ANN)-assisted PSO[21]. Force matching[16–18,28] has been reformulated as a supervised learning problem in CGNet[25], using ANN and point forces as features to learn the potential of mean force of a polypeptide in water. Automatic learning of both the CG FF and its "functional form" (abstract featurization) was then introduced in CGSchNet[26] using graph convolutional neural networks (GCNN).

Such approaches, relying exclusively on equilibrium AA trajectories for CG FF calibration (exclusively *bottom-up*), are exposed to MD sampling problems and potential inaccuracies in the reference AA FF (which are then transferred into the CG FF). Automatic approaches capable of optimizing the accuracy of the CG models, while at the same time guaranteeing their transferability to different conditions or system variants than those used during FF calibration, would be fundamental for the development of next-generation transferable CG FFs.

Using phosphatidylcholine (PC) lipids as a test case, here we describe an automatic multi-objective optimization approach that allows to develop accurate and transferable CG lipid FFs. We build on *SwarmCG*,[23] a CG FF optimization algorithm based on fuzzy self-tuning PSO[49] (FST-PSO), recently developed to optimize bonded parameters in CG molecular models. We designed a comprehensive general strategy that now allows *SwarmCG* to optimize also the *non-bonded* parameters of a FF in order to improve the accuracy of the CG models. We chose lipids as a test case, as an example of molecular systems for which we can avail of experimental data (*e.g.*, area per-lipid, bilayer thickness, etc.) and of reliable all atom FFs. Here, in a new multi-objective version of *SwarmCG* (https://github.com/GMPavanLab/SwarmCGM), we provide the opportunity to combine *bottom-up* and *top-down* reference information for calibrating CG lipid FFs using simultaneously high-resolution (AA) MD simulations and the experimental data. Using new metrics based on optimal transport (OT)[50] for deriving the *bonded* and *non-bonded* interaction terms of the CG FF, we show that *SwarmCG* can simultaneously and iteratively optimize CG models of different types of PC lipids in parallel, improving the transferability of the optimized CG parameters among different lipid types, and also to those which are not included in the training set. Several FF calibration experiments demonstrate that this multi-objective approach can be successfully applied for generating new and custom lipid FFs across different resolutions. Furthermore, a stress-test of *SwarmCG* against a state-of-the-art CG lipid FF (Martini 3.0)[24] proves the robustness of the software.

## II. METHODOLOGY

### A. Optimization strategy

Our multi-objective CG FF optimization strategy relies on the complementary use of structure-based information from high-resolution molecular simulations (*bottom-up:* AA MD), providing knowledge on the submolecular structure and dynamics of the systems, and of experimental data (*top-down*: *e.g.*, area per lipid, bilayer thickness), used to guide the calibration of the models on a larger scale (Fig. 1a). Similar multi-objective strategies, based on the simultaneous combination of simulation and experimental data used as the references for FFs fitting, have been employed also by others for the development, *e.g.,* of different types of AA FFs[29,36,51,52]. In our approach, the discrepancies observed between the data provided by the CG models and the *bottom-up* and *top-down* reference data are measured *via* a global scoring function (Eq. 1) and minimized through an iterative optimization procedure. Executed in parallel using in the training set multiple bilayers, each composed of a different type of lipid, our procedure can output optimal versions of CG lipid FFs, offering optimal consistency with the experimental data set as target. Importantly, the quality and completeness of the information embedded in the training set (*e.g.*, number of different types of lipids, number of different temperatures used, resolution and topology of the CG representations, accuracy of the AA MD simulations and of the experimental data set used as the targets) directly conditions the accuracy of the CG FFs optimized *via SwarmCG*, as well as their capacity to transfer to other types of lipids (not included in the training set).

For our demonstrations, here we use in the training sets up to 5 PC lipids that span a range of different tail characteristics (length, unsaturation, and combinations of those), for which accurate experimental measurements for the area per lipid (APL) and phosphate-to-phosphate bilayer thickness ($D_{HH}$) are available from lamellar bilayer isolates in the liquid phase[53,54] (Fig. 1b, Table S2). As proof of concept of this method for optimizing CG FFs for lipid models with different resolutions, we first focus on parametrizing custom CG models where the lipid molecules are represented at low resolution using either 5 beads per lipid (lowest resolution), or 6-8 beads per lipid (slightly higher resolution). We then test the same approach on high resolution PC lipid models (3-5 heavy atoms per bead). In such a case, as a stress-case for the method, we start from the state-of-the-art version of the widely used Martini FF (*i.e.*, Martini 3.0)[24]. Given the notable level of usage and testing of this CG FF (especially for the simulation of lipids), we use this as a control case to check that *SwarmCG* does not produce CG parameters deviating much from a FF that is already quite evolute in terms of accuracy and reliability. All the models in these demonstrations make use of the same simple and computationally-efficient FF functional form (*i.e.*, the standard one used in Martini[24] lipid models, see also Sec. S1), with *bonded* interactions described by harmonic terms for bonds and angles, and *non-bonded* interactions described by Lennard-Jones (LJ) and Coulomb potentials. The parameters of the CG FFs are iteratively optimized using FST-PSO[49] (one of the most efficient PSO variant to date[55]) and by running at each iteration 200 ns of CG MD simulation of lipid bilayer patches composed of 128 lipids (which preliminary tests indicated being enough to reach successfully the MD equilibrium and sufficient sampling in the simulated CG bilayer systems), from which the scores of the CG models are measured and used for improving the FFs accuracy according to a loss function.

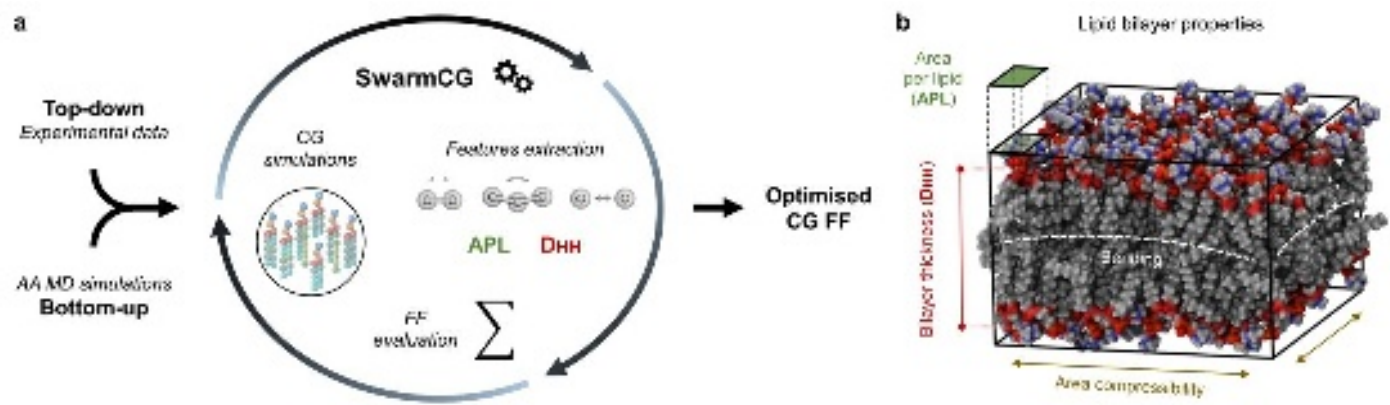

**Figure 1.** Multi-objective optimization process implemented in *SwarmCG*. (a) *SwarmCG* simultaneously relies on *bottom-up* and *top-down* references to iteratively optimize CG lipid FFs using higher-resolution AA MD simulations and experimental data. (b) Properties of lipid bilayers (APL and $D_{HH}$) used for calculating the *top-down* component of the loss function.

In classic PSO[48], a swarm of individuals (referred to as "particles", each representing a set of parameters to be optimized) moves iteratively inside a bounded multidimensional search space and cooperates to identify the best solution for a problem according to a loss function, without using analytical gradients. Settings referred to as "social" and "cognitive" attraction respectively favor the collaboration among particles and their tendency to rely on individual experience. In FST-PSO[49], faster convergence is achieved with the introduction of fuzzy logic for adjusting attraction settings independently for each particle and dynamically during optimization. This approach is particularly competitive in computationally expensive black-box optimization problems and effectively handles noisy data. Applied to our CG FF calibration problem, this enables using a variational principle for high-dimensional FF parameterization with limited concern over the impact of the noise originating from MD sampling. In our demonstrations, the loss function evaluates the CG FF parameters of up to 5 different lipids (Fig. 2a,b) simultaneously (*i.e.*, each particle of the swarm is tasked with running up to 5 CG MD simulations). The optimization problem is thus formulated for maximizing the thermodynamic consistency of the optimized CG FF for all tested PC lipids (potentially simulated using different temperatures), increasing the sampling of the data while at the same time limiting the number of local minima on the loss surface. Therefore, the computational cost of running multiple simulations for calculating the relevance of a single set of FF parameters (*i.e.*, a particle of the swarm) is compensated by an even faster convergence, enabled by an information-rich loss function. Because swarm optimization already constitutes an embarrassingly parallel workload, and each particle is tasked with running independent MD simulations (*i.e.*, another layer of parallelization), this approach also efficiently leverages high-performance computing (HPC) resources (see Sec. S2).

### B. Loss function and optimal-transport-based metrics

#### 1. *Top-down* components

Our process is based on the minimization of a loss function encompassing the distances from the *bottom-up* and *top-down* target objectives. Our construction of the loss function aims at: (i) reducing a many-objective optimization problem to a single-objective one (global FF accuracy score); and (ii) using as priority reference the available experimental data (APL and $D_{HH}$), while the features calculated from AA MD simulations (which may suffer from sampling issues or FF inaccuracies) are used only as guidance during optimization, and for restricting the number of possible solutions to a given optimization problem. The loss function takes the form

$$loss = \sqrt{\frac{\Delta APL_{global}^2 + \Delta D_{HH\,global}^2 + OT\text{-}B_{global}^2 + OT\text{-}NB_{global}^2}{4}}, \quad (1)$$

where $\Delta APL_{global}$ and $\Delta D_{HH\,global}$ are the aggregated APL and $D_{HH}$ deviations with respect to experimental values, calculated across all CG lipid bilayers used for an optimization (loss components 1 and 2) as

$$\Delta APL_{global} = \sqrt{\frac{\sum(w_1 + \min(\max(0, \Delta APL_{lt} - E_{tol}), \Delta APL_{cap}))^2}{L_g}}, \quad (2)$$

and

$$\Delta D_{HH\,global} = \sqrt{\frac{\sum(w_1 + \min(\max(0, \Delta D_{HH\,lt} - E_{tol}), \Delta D_{HH\,cap}))^2}{L_g}}, \quad (3)$$

where $\Delta APL_{lt}$ and $\Delta D_{HH\,lt}$ are the APL and $D_{HH}$ absolute % deviations with respect to the experimental values for a bilayer composed of lipid type $l$ simulated at temperature $t$, $E_{tol}$ is set to 1.5 and represents the tolerated error in $\Delta APL_{lt}$ and $\Delta D_{HH\,lt}$ accounting for experimental error in measurements, $w_1$ is a weight (empirically set to 10) that prioritizes using the target experimental data over the AA reference data during the FF optimization, $L_g$ is the number of different lipid bilayer simulations used in the training set (a bilayer of lipid type $l$ can be simulated at different temperatures), and $\Delta APL_{cap}$ and $\Delta D_{HH\,cap}$ are set to 30 and used to cap $\Delta APL_{lt}$ and $\Delta D_{HH\,lt}$ values for limiting noise during the first steps of an optimization procedure, by allowing to disregard uninformative loss values that can be produced when putative CG FF parameters would induce a disassembly or explosion of the bilayer during MD simulations. By capping $\Delta APL_{global}$ and $\Delta D_{HH\,global}$ values, then $OT\text{-}B_{global}$ and $OT\text{-}NB_{global}$ (Eq. 8, 10) are able to guide the optimization even in otherwise potentially uninformative conditions. We define the convergence criterion as 10 swarm iterations without improving loss.

### 2. Optimal-transport-based metrics

Regarding the *bottom-up* component, as preliminary steps we obtain well-sampled equilibrium AA MD trajectories of lipid bilayers to be used as references for each lipid to be used in the training sets (see Sec. S3), and map the AA lipid models at the desired CG resolution (Fig. 2a). The AA-to-CG mapping determines the chemical identity/correspondence of each CG bead, bond and angle, defining also the number of parameters to be optimized in the CG FF (Figs. 2c, additional details provided in Sec. S4). Reference bond and angle distributions, as well as the distance distributions between each type of particle (within a 25 Å cut-off), are computed from each AA-mapped MD trajectory, and compared to those calculated using the corresponding CG models at each iteration during optimization (inexpensive via MDAnalysis[56,57]). Altogether, the discrepancies between such average AA and CG quantities measure how closely a putative CG FF matches the AA description of the molecular systems[18].

For evaluating the mismatch between corresponding AA *vs.* CG bond and angle distributions, we employ the OT-based Wasserstein distance[58,59] (a.k.a., Earth Movers' Distance, EMD) with an underlying symmetric and positive-definite distance matrix (hereafter referred to as "OT-B metrics", Fig. 2d). This metrics has been already proven well-suited for parametrizing the *bonded* terms of CG models of complex and flexible molecules in a previous version of *SwarmCG*[23]. Noteworthy, OT-based metrics offer several interesting features: (i) multimodal distributions are properly handled; (ii) distances are robust to noise; (iii) distances are quantified in interpretable units (*e.g.*, Å, degrees); and (iv) their computations are inexpensive. In particular, here we introduce a new metrics that relies on OT for comparing the spatial distribution of particles in equilibrium MD trajectories (hereafter referred to as "OT-NB metrics", Fig. 2e). The OT-NB metrics employs the Wasserstein distance[58,59] on the distance distribution between particles, with an underlying distance matrix accounting for the differences in between radial shell volumes. This metrics can be considered as an OT-based adaptation of the Kirkwood-Buff integrals[60], which is particularly well-suited for quantifying differences in the spatial organization of particles in molecular systems described at different resolutions (*e.g.*, AA *vs.* CG).

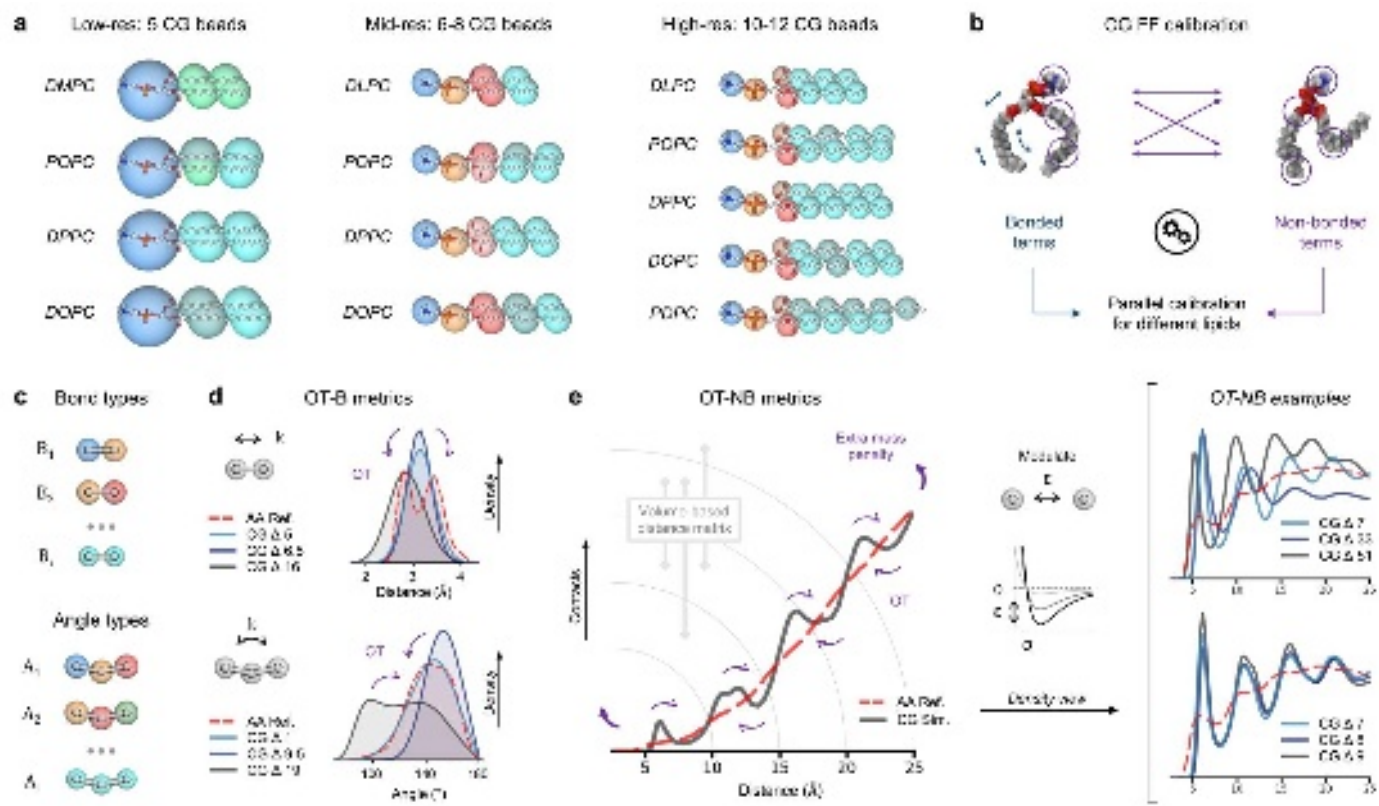

**Figure 2.** Description of the CG representations applied and the metrics used in this modified version of *SwarmCG*. (a) AA *vs.* CG mapping for the benchmark lipids used in the training set for optimization at different resolutions: 1,2-dilauroyl-sn-glycero-3-phosphocholine (DLPC), 1,2-dimyristoyl-sn-glycero-3-phosphocholine (DMPC), 1,2-dipalmitoyl-sn-glycero-3-phosphocholine (DPPC), 1-palmitoyl-2-oleoyl-glycero-3-phosphocholine (POPC), 1,2-dioleoyl-sn-glycero-3-phosphocholine Δ9-Cis (DOPC) and 1-palmitoyl-2-docosahexaenoyl-sn-glycero-3-phosphocholine Δ4,7,10,13,16,19-Cis (PDPC, 16:0-22:6). (b) *Bottom-up* components of the scoring function: *SwarmCG* optimizes in concert the *bonded* and *non-bonded* terms of a CG FF, iterating CG MD simulations of bilayers composed of different types of lipids. (c) CG bonds and angles are classified according to the CG beads involved, and are attributed specific parameters. (d) Principle of the OT-B metrics used for structure-based information related to bonded FF terms. (e) Principle of the OT-NB metrics used for structure-based information related to non-bonded FF terms.

Given 2 histograms $P, Q$ the EMD as initially defined by Rubner et al.[59] is:

$$EMD(P,Q) = \frac{\min_{\{f_{ij}\}} \sum_{i,j} f_{ij} d_{ij}}{\sum_{i,j} f_{ij}}, \quad (4)$$

with $f_{ij} \geq 0$, $\sum_j f_{ij} \leq P_i$, $\sum_i f_{ij} \leq Q_j$ and $\sum_{i,j} f_{ij} = \min(\sum_i P_i, \sum_j Q_j)$, where $\{f_{ij}\}$ represents the optimal transport plan, each $f_{ij}$ represents the amount transported from the $i$ supply bin to the $j$ demand bin, and $d_{ij}$ is the distance matrix between bin $i$ and bin $j$ in the histograms. In this study, the domains of the bond and angle distributions $D_b$ and $D_a$ are set to [0, 50] Å and [0, 180] degrees, respectively, for the distributions of all AA-mapped and CG bond and angle, with histogram bandwidths set to 0.1 Å and 2 degrees. The domain of all the AA-mapped and CG distance distributions between pairs of bead types $D_p$ is set to [0, 15] Å and the bandwidth used is 0.25 Å. All EMD calculations used in this study rely on the implementation of PyEMD[58,61].

To calculate the EMD between 2 corresponding AA-mapped vs. CG bonds or angles, noted $OT\text{-}B_{bond(i,l)}$ and $OT\text{-}B_{angle(i,l)}$ in the next section, we use normalized histograms ($\sum_i P'_i = 1$ and $\sum_j Q'_j = 1$) and a symmetric positive-definite distance matrix. For obtaining the EMD between 2 corresponding AA-mapped vs. CG distance distributions between pairs of bead types, noted $OT\text{-}NB_{bead\ pair(l)}$ in the next section, we first normalize the bins in each of the 2 histograms so that $\sum_i P'_i + \sum_j Q'_j = 2$, with $P'_i = \frac{2 P_i}{\sum_i P_i + \sum_j Q_j}$ and $Q'_j = \frac{2 Q_j}{\sum_i P_i + \sum_j Q_j}$. We then define a distance matrix that accounts for the differences in between radial shell volumes as:

$$d_{ij} = \begin{cases} i < j, & V_j/V_i \\ i = j, & 0 \\ i > j, & V_i/V_j \end{cases}, \quad (5)$$

with $V_i$ and $V_j$ the volumes of the radial shells for bin $i$ and $j$. The optimal transport plan $\{f_{ij}\}$ calculated here consists in a partial matching of the 2 compared histograms[58,61] because of the normalization we apply. We finally account for the extra or missing mass that is left out of $\{f_{ij}\}$ using:

$$OT\text{-}NB_{bead\ pair(l)} = \left(EMD(P',Q') + \left(\frac{\max(\sum_i P_i, \sum_j Q_j)}{\min(\sum_i P_i, \sum_j Q_j)} - 1\right)\right) * 100. \quad (5)$$

Because the distance matrices $d_{ij}$ are symmetric, the OT-B and OT-NB distances inherits the properties of metrics[58,61].

### 3. *Bottom-up* components

Finally, for the definition of our global loss function (Eq. 1), we aggregate the OT-B and OT-NB distances obtained from the comparison between a set of CG MD simulations (*i.e.*, a particle of the swarm) and their corresponding reference AA MD trajectories (*bottom-up*), together with the discrepancies observed between CG *vs.* experimental APL and D$_{HH}$ measurements (*top-down*). The OT-B distances from the reference AA MD trajectories are calculated as:

$$OT-B_{bond\ type} = \sqrt{\frac{\sum(w_2 \times OT-B_{bond(i,l,t)})^2}{B_{i,l,t}}}, \quad (6)$$

where $OT-B_{bond\ type}$ quantifies the deviation of the CG models from the reference AA trajectories in terms of bond distributions for a given bond type, $B_{i,l,t}$ is the number of instances of this bond type in CG model topologies across all simulations used in an optimization, $OT-B_{bond(i,l,t)}$ is the OT-B distance for each instance of this bond type across all simulations used in an optimization (a bond of a given type can be present multiple times in the topology of a single lipid model), $w_2$ is a weight that prioritizes minimizing the OT-B distances of the bonds over those of the angles (see below); and

$$OT-B_{angle\ type} = \sqrt{\frac{\sum(OT-B_{angle(i,l,t)})^2}{A_{i,l,t}}}, \quad (7)$$

where $OT-B_{angle\ type}$ quantifies the deviation of the CG models from the reference AA trajectories in terms of angle distributions for a given angle type, $A_{i,l,t}$ is the number of instances of this angle type in CG model topologies across all simulations used in an optimization, $OT-B_{angle(i,l,t)}$ is the OT-B distance for each instance of this angle type across all simulations used in an optimization (an angle of a given type can be present multiple times in the topology of a single lipid model); and

$$OT-B_{global} = \sqrt{\frac{\sum OT-B_{bond\ type}^2 + \sum OT-B_{angle\ type}^2}{B_g + A_g}}, \quad (8)$$

where $OT-B_{global}$ is the global OT-B deviation score for the CG FF being optimized (loss component 3), $B_g$ is the number of different bond types in this FF and $A_g$ the number of different angle types in this FF. The OT-NB distances from the reference AA MD trajectories are calculated as:

$$OT-NB_{bead\ pair} = \sqrt{\frac{\sum(OT-NB_{bead\ pair(l)})^2}{P_l}}, \quad (9)$$

where $OT-NB_{bead\ pair}$ quantifies the deviation of the CG models from the reference AA trajectories in terms of distance distributions between pairs of bead types, $OT-NB_{bead\ pair(l)}$ is the OT-NB distance for this pair of bead types calculated from each CG MD simulation in which this interaction is sampled, $P_l$ is the number of instances of this pair of bead types across all lipids used in an optimization; and

$$OT-NB_{global} = \sqrt{\frac{\sum OT-NB_{bead\ pair}^2}{P_g}}, \quad (10)$$

where $OT\text{-}NB_{global}$ is the global OT-NB deviation score for the CG FF being optimized (loss component 4), and $P_g$ is the number of different pair types in this FF.

Therefore, the components of the loss function are empirically weighted according to only 2 parameters: (i) $w_1$ slightly prioritizes minimizing the APL and D$_{HH}$ discrepancies over the OT-B and OT-NB distances, which regulates the extent to which structure-based information is discarded for better fitting experimental measurements, and (ii) $w_2$ allows to obtain comparable OT-B metrics for the bond and angle deviations and is set to 50, meaning that an OT-B of 0.4 Å between corresponding CG *vs.* AA-mapped bond distributions (noted $OT\text{-}B_{bond(i,l)}$) is considered equivalent to an OT-B of 20 degrees between corresponding CG *vs.* AA-mapped angle distributions (noted $OT\text{-}B_{angle(i,l)}$).

**4. Modulation of the loss function for mixed *bottom-up* and *top-down* FF calibration**

The simple form used for the aggregation of the different objectives in the loss function (Eq. 1) enables modularity. In this modified version of *SwarmCG*, if no AA trajectories are provided for a given optimization procedure, the *bottom-up* components of the loss function described in Eq. 1 will be discarded and the loss will automatically become:

$$loss' = \sqrt{\frac{\Delta APL_{global}^2 + \Delta D_{\text{HH}global}^2}{2}}. \tag{11}$$

User-defined configuration files allow complete modularity of the parameters being optimized and of the reference data being used, notably for performing CG lipid FF calibrations in a mixed *bottom-up* and *top-down* fashion. For example, this allows to make use of *bottom-up* reference AA trajectories only for lipids for which an accurate AA FF is available (see Sec. S3 and Table S1), while parameters that are specific to CG lipids for which the AA FFs are still inaccurate to date (*e.g.*, containing highly unsaturated tails, such as SDPC and PDPC[53]) can be calibrated exclusively with respect to *top-down* experimental data. In this case, these parameters can be evaluated in parallel and in context with other FF parameters subjected to both *bottom-up* and *top-down* reference calibration.

## III. RESULTS

### A. Multi-objective automatic optimization of CG models of POPC at different resolutions

As a first demonstration of this approach, we use *SwarmCG* to perform individual multi-objective (*bottom-up* plus *top-down*) optimizations of 3 different-resolution CG models of POPC. In particular, for this first test we use: (I) a low-resolution (5 beads) implicit-solvent CG model, (II) a mid-resolution (8 beads) implicit-solvent CG model, and (III) a high-resolution (12 beads) explicit-solvent CG model of a POPC bilayer composed of 128 lipids (64 per leaflet). These three CG models are then optimized individually, with the prime purpose to check how *SwarmCG* copes with the different resolution of the models.

In the case of (I) and (II), the parametrization of the coarser CG models is performed completely *ab-initio* (Fig. 2a,b). In such cases we initially set all parameters randomly, while in the first particle in the first swarm iteration all LJ ε are set to 4 kJ/mol (it can be a random number, but too low or too high ε values should be avoided in order to prevent all CG MD simulations to crash in the first swarm iteration). All parameters are then iteratively changed by *SwarmCG* with the goal of minimizing the loss function. In particular, the following terms of the CG FFs are calibrated: (i) equilibrium values for bonds and angles; (ii) force constants for bonds and angles (*bonded* terms); and (iii) LJ σ and ε parameters defining all interactions between pairs of CG bead types (*non-bonded* terms). For obtaining the LJ σ parameters, we optimize the values of the radii attributed to each bead type and apply the Lorentz-Berthelot rule[62] (see also Sec. S5 and S6 in the SI). As initial parameters, we use: (i) the average equilibrium values of the bonds and angles computed from AA-mapped MD trajectories; (ii) arbitrary values for all force constants; and (iii) arbitrary values for the bead radii and LJ ε parameters. The green curves in Fig. 3a,b (loss) indicate that the optimizations converged after ∼50 and ∼40 swarm iterations, respectively. Independently of the resolution applied, at convergence the models correctly reproduce the APL and $D_{HH}$ experimental data used as target (Fig. 3a,b: yellow and blue curves converging to reference black line, set to 0). The OT-B and OT-NB distances are effectively minimized (Fig. 3a,b: black and olive curves), indicating that both CG models globally reproduce the structural features present in the reference AA MD trajectories (Figs. S3-4: low-resolution).

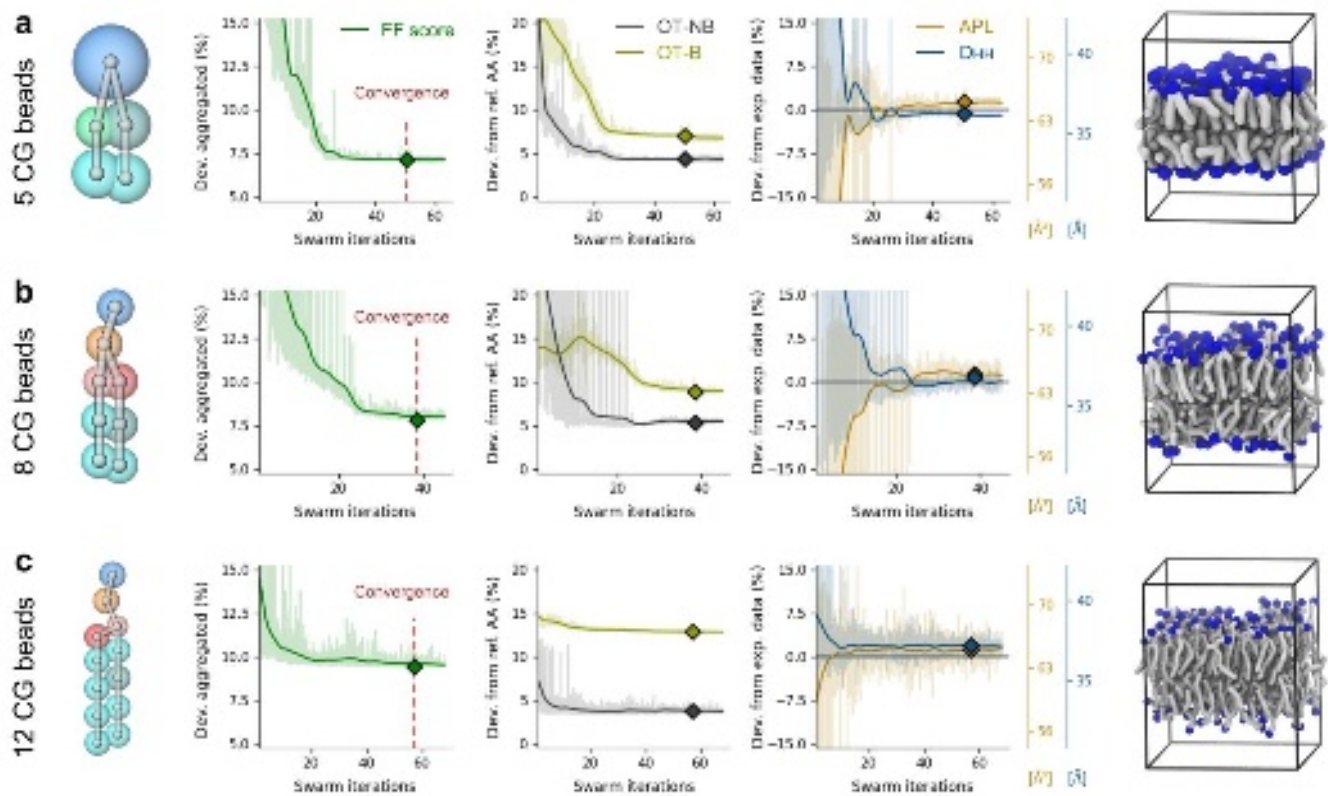

**Figure 3.** Multi-objective optimization of CG FF parameters in models of POPC described at different resolutions, at 303K in the liquid phase. For each modelling resolution: (a) 5 CG beads, (b) 8 CG beads, (c) 12 CG beads – (Left panels) green curve: loss during optimization; (Middle panels) black and olive curves: OT-B and OT-NB metrics for the specific lipid; (Right panels) yellow and blue lines: APL and $D_{HH}$ for the specific lipid model during optimization, displayed with window-averaging (solid) and without (shaded); (Right images) CG MD snapshots of the lipid bilayers for the FF obtained at each resolution. The horizontal black lines set at 0 identify the target experimental APL and $D_{HH}$ values. Diamonds represent values at convergence, obtained with the selected optimized CG FF parameters.

In the case of (III), we test our approach in explicit solvent starting from the current high-resolution Martini 3.0[24] model of POPC (Fig. 2c). The following terms of the CG FF are iteratively optimized: (i) the equilibrium values for bonds and for angles not initially set to 180; (ii) force constants for bonds and angles; and (iii) LJ ε parameters defining all solute-solute interactions between pairs of CG bead types. LJ σ parameters, solute-solvent and solvent-solvent interactions are unchanged. As initial parameters, we use the existing ones in Martini 3.0[24] for this model. The optimization procedure converged after ~60 swarm iterations (Fig. 3c). At convergence the model of POPC correctly reproduces the APL and $D_{HH}$ experimental data used as target (Fig. 3a,b: yellow and blue curves), whereas the original Martini 3.0[24] model produces a small offset (~6 %) on $D_{HH}$ with respect to experimental data. The OT-B distances reached a higher plateau (Fig. 3c: olive curve) than in the 2 previous cases, due to the CG topologies used in Martini 3.0[24] lipid models, which creates small offsets in the CG *vs.* AA-mapped fits of the bond distributions (see Sec. S7). This has no effect on OT-NB distances, which were minimized to a plateau similar to the 2 previous cases (Fig. 3c: black curve).

This first experiment demonstrates the ability of *SwarmCG* to balance *bonded* and *non-bonded* interaction terms to optimize the CG lipid models, independently of the resolution used. In all three cases, the obtained CG models for POPC self-assemble into bilayers and allow to simulate vesicle fusion (Fig. 4: *e.g.*, 5 and 8 CG beads models) consistently with previous studies[63,64]. In terms of computational time, the *ab-initio* calibration of the low-resolution POPC model (Fig. 3a, 5 CG beads and 26 FF parameters) required 36 hours (wall-clock time) to reach 60 swarm iterations using 20 particles in the swarm and using 20 CPUs (each CG simulation running on a single CPU, allowing for a complete parallelization of the swarm of particles). The *ab-initio* calibration of the mid-resolution POPC model (Fig. 3b, 8 CG beads and 55 FF parameters) required 2 days to reach 50 swarm iterations using 25 particles in the swarm and requesting 25 CPUs. The optimization starting from the existing Martini 3.0[24] POPC model (Fig. 3c, 12 CG beads and 40 FF parameters) required 8 days to reach 70 swarm iterations using 23 particles in the swarm and requesting 23 CPUs. We underline that while the time and computational cost required for the optimization may seem non-negligible, the benefit makes the process balance favorable, especially considering that the result of *SwarmCG* in this case, as it will be better described in the next section, is an optimized CG FF.

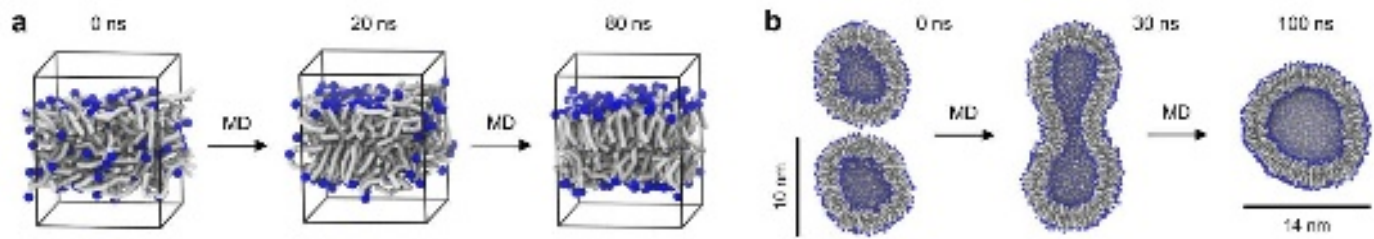

**Figure 4.** Behavior of the POPC models calibrated *ab-initio* in different configurations of unbiased MD simulations in implicit solvent. (a) CG MD snapshots of the low-resolution 8 beads POPC models undergoing self-assembly into a bilayer starting from a dispersed configuration (128 lipids, 310K). (b) CG MD snapshots of the mid-resolution 5 beads POPC models undergoing vesicular fusion starting from 2 pre-assembled vesicles (1196 lipids each, 310K).

Such individual optimizations demonstrate that, from a technical point of view, *SwarmCG* can produce CG models fitting with the experimental data. However, the following comments on the robustness of the results are necessary. As said above, there is no guarantee that optimized CG parameters obtained in such a way are transferable to simulate other lipid systems (*vice versa*, typically, they do not). Furthermore, the individual optimization of multiple parameters (*e.g.*, all *non-bonded* LJ terms) to fit, *e.g.*, the APL of POPC, could lead to potential artifacts. In fact, automatic optimization approaches, such as this one, cope badly with resolving a high-dimensional problem using low-dimensional criteria. Typically, the method can find multiple combinations of LJ parameters that allow reaching the APL target. While the use of multiple references (*bottom-up* and *top-down*) may to some extent alleviate such limitation, here we designed a different approach that imparts not only robustness, but also transferability to the results of *SwarmCG* in this sense (see next sections).

### B. Parallel *ab-initio* calibration of low-resolution implicit-solvent CG lipid force fields

In order to provide more information for guiding the optimization, here we introduce an additional transferability constraint. This consist in the parallel optimization of multiple lipid bilayer systems of different types, which are optimized as in the previous section in an iterative way against *bottom-up* and *top-down* references. We conducted this test using, *e.g.*, POPC, DMPC, DOPC and DPPC lipid bilayers (Fig. 5a). The difference from the previous experiment is that in this case, at every iteration, the parameters change affects all CG beads of the same type in all 4 lipid systems in the same way. Furthermore, the discrepancies between all 4 systems and their *bottom-up* and *top-down* references become all part of the scoring function which is iteratively minimized by *SwarmCG*. This introduces an automatic transferability constraint, as the number of parameters combinations that can minimize the distance from all objectives in all cases are strongly reduced. In a sense, this is an endogenous constraint, in that the automatic optimization self-regulates under the condition that this has to satisfy and optimize all systems (and not only one, POPC, as in the previous tests of Fig. 3).

In this section, we pursue our demonstration by running a completely *ab-initio* optimization of an implicit-solvent PC lipid CG FF using 4 different low-resolution models (5 CG beads) in the training set using an analogous combined *bottom-up* and *top-down* optimization approach as described in the previous section. At every iteration, *SwarmCG* attempts to minimize the distance between the 5 beads per lipid CG models of POPC, DMPC, DOPC and DPPC lipid bilayers (each composed of 128 lipids), their experimental APL and $D_{HH}$ data (*top-down*), and the *bottom-up* information extracted from their respective AA MD equilibrated trajectories. The conditions of the four lipid models parallel optimization conducted in this experiment are identical to these of the low-resolution experiment in Sec. III A, for the types of parameters being optimized in the FF and the initialization of the swarm of particles. After ~100 swarm iterations this parallel procedure reaches convergence (Fig. 5a: green curves), identifying a set of CG parameters allowing to approach closely the experimental APL and $D_{HH}$ values set as the target for all the 4 lipids in the dataset (Fig. 5a, right: yellow and blue curves identifying the percentage deviation from the experimental APL and $D_{HH}$ targets, identified by the black lines set to 0). The OT-B and OT-NB distances are successfully minimized (Fig. 5a: black and olive), indicating that the structural features described by the AA trajectories are being reproduced in the CG models. Although the lipids are here represented at low-resolution (5 CG beads each) and the functional form of the FF being applied is simple (LJ and harmonic potentials), our procedure still allows to identify a satisfying set of FF parameters which are intrinsically transferable among all 4 lipids in the training set (they are optimized under such constraint), and which guarantee good correspondence with the experiments in all cases. This *ab-initio* FF calibration required 8 days of computation to reach 100 swarm iterations using 23 particles and requesting 46 CPUs (42 FF parameters, 4 CG simulations of small bilayer patches per particle of the swarm).

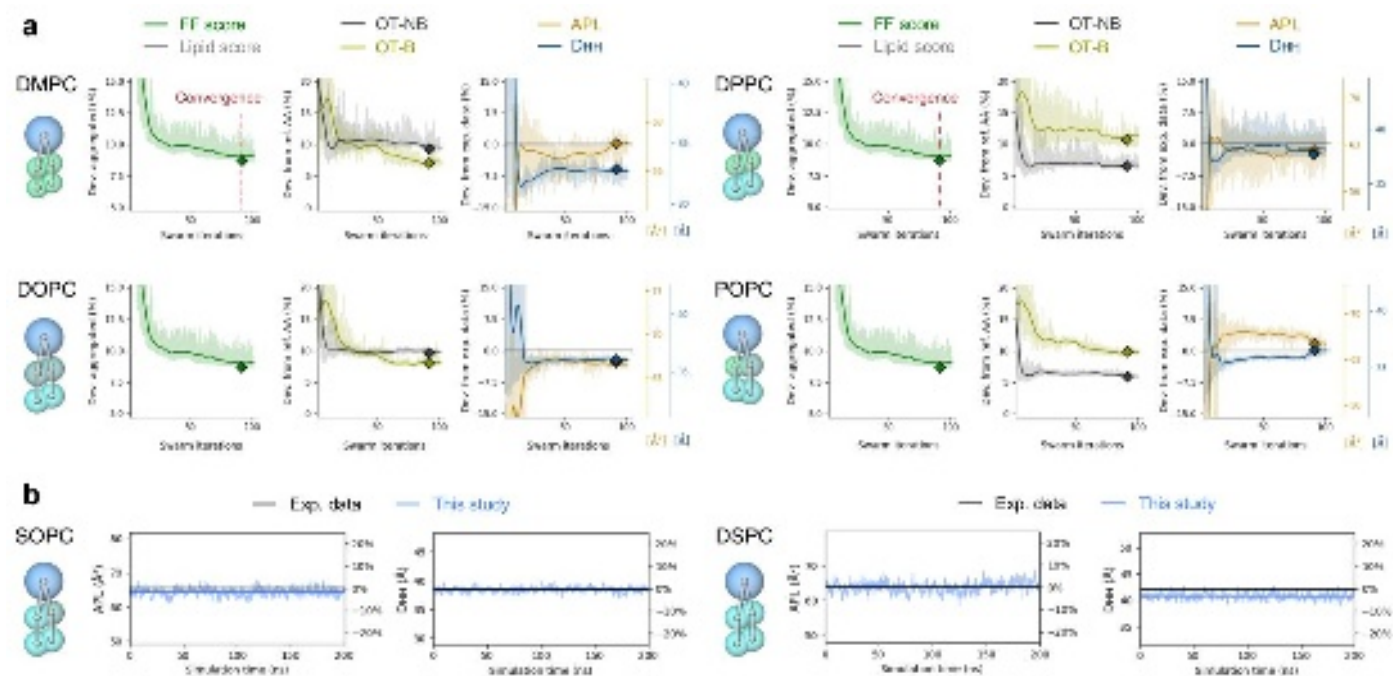

**Figure 5.** Multi-objective *ab-initio* calibration of CG FF parameters using 4 low-resolution models of PC lipids in the training set, in the liquid phase and in implicit solvent (DMPC: 303K, DPPC: 323K, DOPC: 303K, POPC: 303K). (a) For each lipid – (Left panels) green curve: loss during optimization, grey curve: loss calculated per lipid; (Middle panels) black and olive curves: OT-B and OT-NB metrics for the specific lipid; (Right panels) yellow and blue lines:

APL and D$_{HH}$ for the specific lipid during optimization, displayed with window-averaging (solid) and without (shaded). The horizontal black lines set at 0 identify the target experimental APL and D$_{HH}$ values. Diamonds represent values at convergence, obtained with the selected optimized CG FF parameters. (b) APL and D$_{HH}$ measured across 200 ns for SOPC (303K) and DSPC (333K) using transferred FF parameters.

Considering that the obtained CG FF parameters are transferable between 4 different PC lipids, a relevant question is to what extent these are transferable also to other PC lipids which are not in the training set, and which structure can be described by combinations of the same CG beads (and by their *bonded* parameters). To tackle such a question, starting from the optimized CG FF parameters obtained at convergence with *SwarmCG* in Fig. 5a, we apply these parameters *a posteriori* to parametrize 2 additional low-resolution models of SOPC and DSPC lipids, and simulate the corresponding bilayers composed of 128 lipids in the liquid phase, at 303K and 333K respectively. It is worth noting that these models can be assembled using the previously calibrated CG beads as building blocks (*non-bonded* parameters), but most of the bonded parameters are still unknown. According to our topology definition, only those of the unsaturated tail of POPC can be recycled (see Sec. S4). This test therefore required another preliminary optimization procedure for calibrating exclusively the 12 unknown *bonded* parameters (3 bonds and 3 angles: equilibrium values and force constants), which we perform this time in a fully *top-down* fashion (no AA trajectories are provided for reference, as described in Sec. II B 4). The calibration of the additional 12 *bonded* parameters required 10 hours of computation to reach 20 swarm iterations using 17 particles and requesting 34 CPUs (2 CG simulations of small bilayer patches per particle of the swarm).

The APL and D$_{HH}$ measured for the obtained CG models of SOPC and DSPC across 200 ns of MD simulation of patches of bilayers of 128 lipids (parametrized using the transferred CG parameters obtained from the optimization of Fig. 5a) are found in very good agreement with experimental data (Fig. 5b: blue data from the CG models *vs.* black experimental target APL and D$_{HH}$ data), with a maximum deviation of <3% in the calculation of the D$_{HH}$ of DSPC. This is a remarkable result, as this second test using the low-resolution models of Fig. 5 demonstrates the ability of our procedure to generate *ab-initio* CG lipid FFs even using incomplete training sets. Namely, the CG parameters obtained by optimizing the FF including some lipids in the training set can be then transferred also to lipid molecules that are outside the training set. It is worth noting that here we used as training set 4 CG simulations of small bilayer patches composed of 4 different types of lipids. In this way, we could further calibrate FF parameters for modelling different types of lipids (*e.g.*, SOPC and DSPC) with a reduced computational effort, by transferring some of the previously calibrated FF parameters, and calculating only the missing parameters. In comparison, obtaining the single POPC model in Fig. 3a required the calibration of 26 FF parameters, while obtaining 4 and then 6 different lipid models required the calibration of only 42 and 54 FF parameters in total, respectively. While this is not the central purpose of this work, this demonstrates how the present approach could be scaled for the calibration of general FFs for many different types of lipids. Importantly, it is not forcefully required to provide AA reference trajectories for each of the lipids for which one desires to optimize CG FF parameters for, as the present approach enables *bottom-up* and/or *top-down* scoring of the FF being calibrated.

### C. Parallel *ab-initio* calibration of mid-resolution implicit-solvent CG lipid force fields

In this section we also demonstrate the *ab-initio* calibration of an implicit-solvent PC lipid CG FF using 4 different lipid types in the training set, while in this case the lipids are represented by mid-resolution CG models (6-8 CG beads per lipid). The combined *bottom-up* and *top-down* multi-objective approach is always the same, and the conditions of the optimization are identical to these of the mid-resolution experiment in Sec. III A, for the types of parameters being optimized in the FF and the initialization of the swarm of particles. Also in this case, *SwarmCG* outputs transferable optimized CG parameters for POPC, DMPC, DOPC and DPPC lipid bilayers matching the experimental APL and $D_{HH}$ data for all 4 lipids within the experimental error (Fig. 6a: yellow and blue curves). The OT-B and OT-NB distances are effectively minimized (Fig. 6a: black and olive curves), with the structural features of the AA trajectories being reproduced in the CG models (Figs. S5-6). This *ab-initio* FF calibration required 40 hours of computation to reach 40 swarm iterations using 27 particles and requesting 54 CPUs (79 FF parameters, 4 CG simulations of small bilayer patches per particle of the swarm).

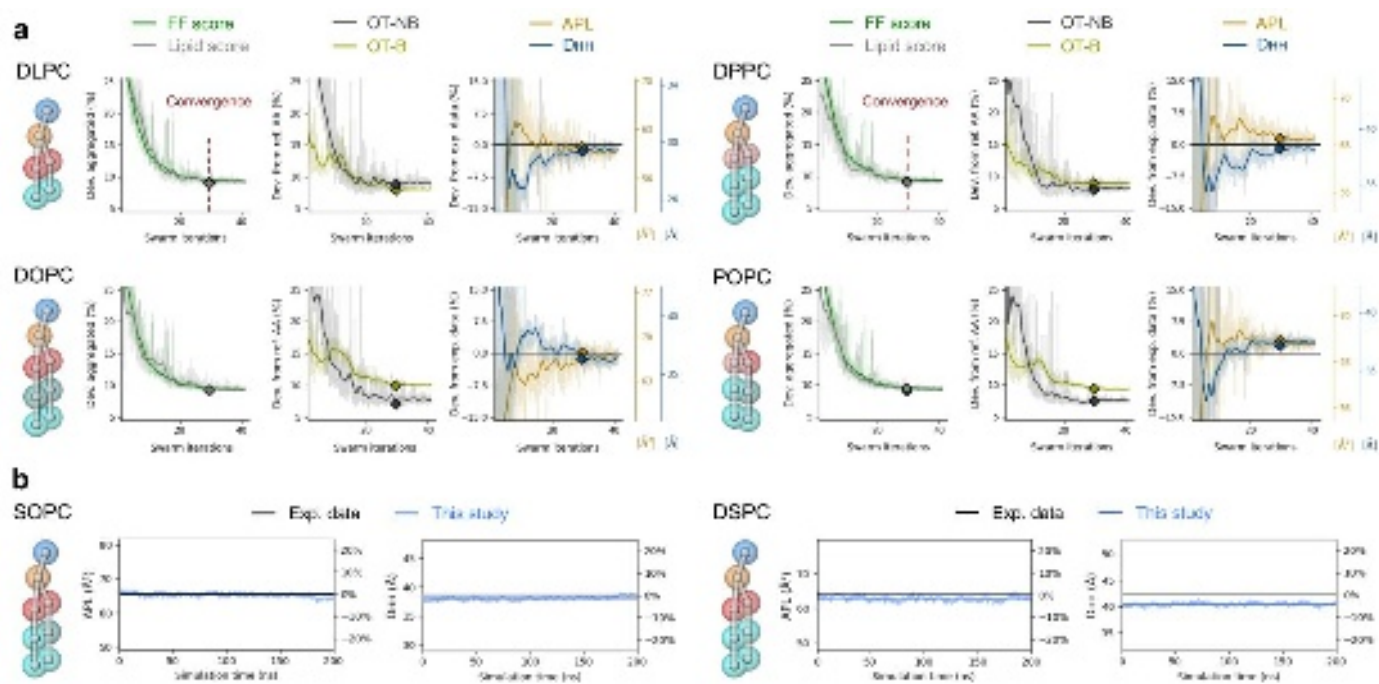

**Figure 6.** Multi-objective *ab-initio* calibration of CG FF parameters using 4 mid-resolution models of PC lipids in the training set, in the liquid phase and in implicit solvent (DLPC: 303K, DPPC: 323K, DOPC: 303K, POPC: 303K). (a) For each lipid – (Left panels) green curve: loss during optimization, grey curve: loss calculated per lipid; (Middle panels) black and olive curves: OT-B and OT-NB metrics for the specific lipid; (Right panels) yellow and blue lines: APL and $D_{HH}$ for the specific lipid during optimization, displayed with window-averaging (solid) and without (shaded). The horizontal black lines set at 0 identify target experimental APL and $D_{HH}$ values. Diamonds represent values at convergence, obtained with the selected optimized CG FF parameters. (b) APL and $D_{HH}$ measured across 200 ns for SOPC (303K) and DSPC (333K) using transferred FF parameters.

Also in this test, we evaluate the transferability of the optimized FF parameters obtained at convergence by building mid-resolution models of SOPC and DSPC and simulating bilayers composed of 128 lipids in the liquid phase, at 303K and 333K respectively. In this case, this requires calibrating just one additional angle (equilibrium value and force constant), which we did again in a fully *top-down* fashion. The APL and $D_{HH}$ measured for the obtained models of SOPC and DSPC using transferred FF parameters are found again in very good agreement with experimental data (Fig. 6b), and results are overall comparable to those of the 5 beads per-lipid model. The calibration of the 2 additional *bonded* parameters required 3 hours of computation to reach 15 swarm iterations using 3 particles and requesting 16 CPUs (2 CG simulations of small bilayer patches per particle of the swarm).

Altogether, the results presented in this and in the previous section demonstrates the effectiveness of *SwarmCG*, used with the parallel multi-objective paradigm presented herein, for producing optimized CG lipid FF of variable resolution. The input required for the process are: (i) a pre-defined CG mapping, which defines the resolution in the description of the lipids in the system (in this sense, a too low resolution – *e.g.*, below 5 beads per lipid, was found to produce poor results and inefficiency in reaching the objectives), (ii) available experimental data and (iii) reliable higher-resolution for the systems that one wants to model. As anticipated, the computational cost for the process, which may seem non-negligible, is fully compensated by the fact that, in a few days, it is possible to obtain fairly accurate and transferable CG FFs. In this sense, we point out that the information present in the training set is important. For example, in the tests described above we obtain optimized CG FF parameters that can be transferred among PC lipids (and in particular, for those lipids which share the lipid types optimized herein). More complete FFs can be for sure obtained by including additional systems in the training set (*e.g.*, PA, PG lipids, etc.). In such a case, the cost/benefit balance of the process is of prima importance. Noteworthy, the fact that *SwarmCG* is demonstrated to produce good results already with a minimal training set is very promising, as this could open the possibility to enrich the obtained FFs adding a limited number of systems to the training set. While future employment and testing of *SwarmCG* to develop new FFs for a growing number of classes of molecular systems will provide insights onto the limits and opportunities in terms of obtaining reliable FFs even using incomplete information, the results reported herein aim at demonstrating the potential of the approach, and to prove the robustness of the parameters that can be obtained with *SwarmCG*. An important point to guarantee is that such powerful data-driven approaches do better, comparably, or at least not worse than humans. With this aim, we designed the last control *in silico* experiment described in the following section.

**D. Control test: using *SwarmCG* starting from the Martini lipid force field parameters**

As a last test case, to control the robustness of the approach, we applied the same approach using the current version of the (explicit solvent) lipid models available in Martini 3.0[24]. This time, we used 5 different high-resolution PC lipids to be included in the training set, which optimization is again iteratively conducted in parallel using the same approach described previously for the other cases. Herein, we use this FF as a last control case. In fact, (i) this FF is considered a reference for the simulation of lipids – we thus expect *SwarmCG* to deviate minimally from the Martini 3.0 CG FF parameters –, and (ii) as the Martini FF provides a finer description of the lipids than that of the models developed in the previous sections, offering an additional test case for our approach.

The conditions of the optimization are identical to these of the high-resolution experiment in Sec. III A, for the types of parameters being optimized in the FF and the initialization of the swarm of particles. After ~30 swarm iterations the procedure reaches convergence (Fig. 7a: green curves), identifying a set of FF parameters that moderately improves the overall matching with experimental APL and $D_{HH}$ values set as target for the 5 lipids in the dataset (Fig. 7a: yellow and blue curves). This FF optimization required 7 days of computation to reach 30 swarm iterations using 23 particles and requesting 60 CPUs (42 FF parameters, 5 CG simulations of small bilayer patches per particle of the swarm). We point out that here we are not aiming to provide an updated version of Martini 3.0 for lipids. A richer training set would be needed to this end. Furthermore, one should guarantee that the changes in the optimized LJ ε parameters obtained for the beads do not produce an impairment in the partition free-energies of the Martini beads (which have been developed based on such concept). However, the results of such test encourage us on the robustness of *SwarmCG*. For example, in all 5 cases we obtain results that variate just by some percentage point compared to the results obtained from the same 5 lipids with Martini 3.0. The result of Fig. 7a demonstrates that while in some cases we obtain a slight improvement (*e.g.*, PDPC, POPC and DPPC), the optimized CG parameters perform slightly worse than Martini 3.0 parameters for other ones (*e.g.*, DOPC and DMPC). This shows that in this case *SwarmCG* struggles in finding a way to improve these parameters, and this is as expected, considered the evolute version of this FF. This demonstrates that *SwarmCG* is found robust in this control experiment. The fact that the software cannot change much FFs, such as Martini 3.0, which are already quite optimized, provides even more value to the results obtained in the tests discussed in the two previous sections.

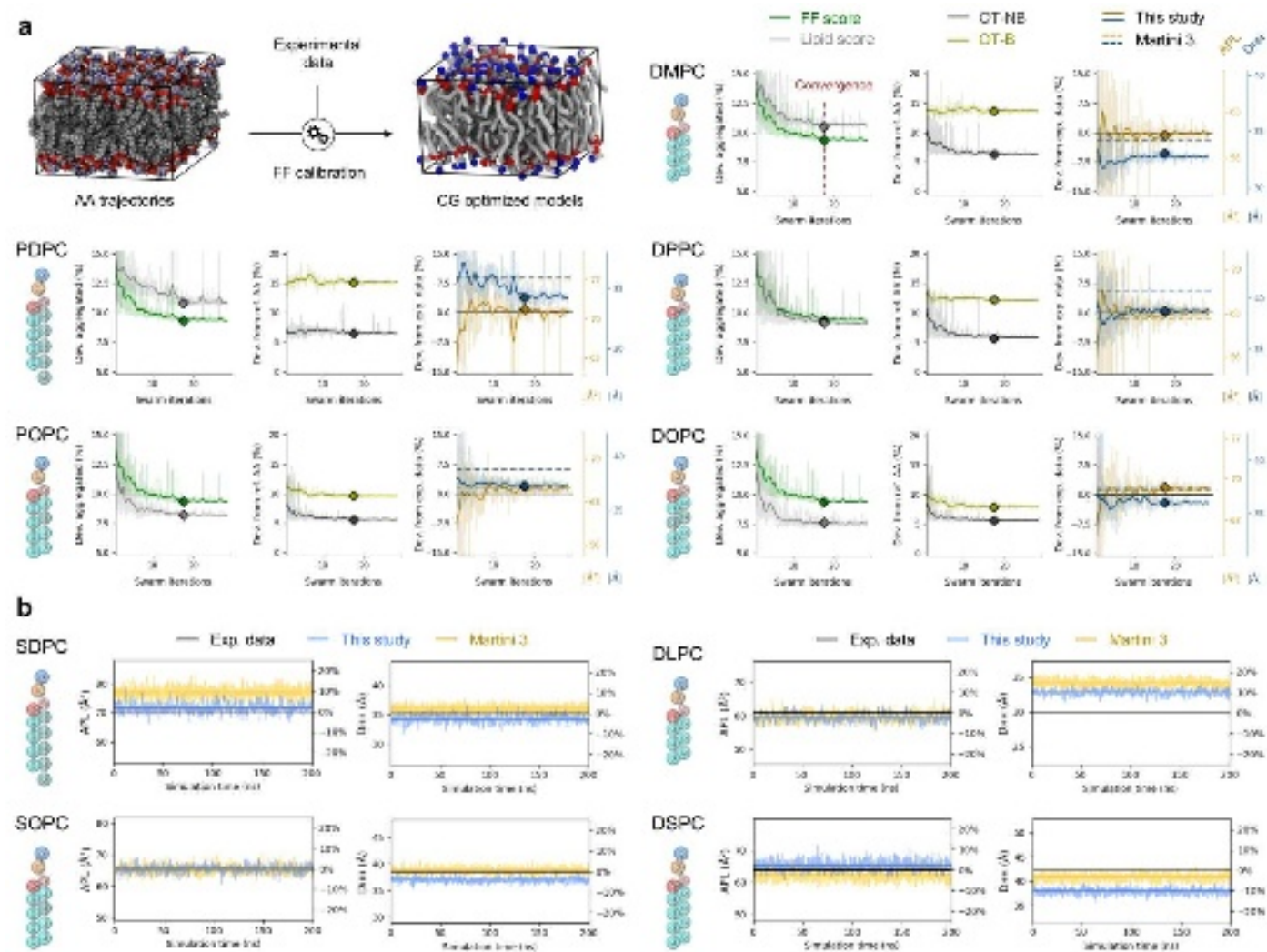

**Figure 7.** Multi-objective optimization of CG FF parameters using 5 high-resolution models of PC lipids in explicit solvent in the training set (DMPC: 303K, DPPC: 323K, DOPC: 303K, POPC: 303K, PDPC: 303K). (a) For each lipid – (Left panels) green curve: loss during optimization, grey curve: CG FF score for the specific lipid; (Middle panels) black and olive curves: OT-B and OT-NB metrics for the specific lipid; (Right panels) yellow and blue lines: APL and $D_{HH}$ for the specific lipid during optimization, displayed with window averaging (solid) and without (shaded), and for Martini models (dotted). The horizontal black lines set at 0 identify target experimental APL and $D_{HH}$ values. Diamonds represent values at convergence, obtained with the selected optimized CG FF parameters. (b) APL and $D_{HH}$ measured across 200 ns for SDPC (303K), DLPC (303K), SOPC (303K) and DSPC (333K) using transferred FF parameters (blue) and for Martini models (yellow).

Also after such a control test, we tried to transfer the obtained FF parameters for modelling DLPC, DSPC, SOPC and SDPC and simulate bilayers in the liquid phase (Fig. 7b), staying within the framework of Martini 3.0[24] lipid models. A known limit with the current CG representations employed in Martini is that they do not allow differentiating between some types of lipids[27,65]. Here, the same models are used for representing, enumerated by pairs: DLPC and DMPC, SDPC and PDPC, DPPC and DSPC, and POPC and SOPC (Table S2). Therefore, while at convergence the experimental APL and $D_{HH}$ values were better fitted for the models of DPPC and POPC (Fig. 7a), the deviation from experimental data is increased when considering the same models for the representation of DSPC and SOPC, respectively (Fig. 7b). Similarly, the model obtained at convergence for DMPC slightly decreased the agreement with experimental $D_{HH}$ value (Fig. 7a), with respect to the original Martini 3.0 model, while the optimized model applied for modeling DLPC yielded better agreement for this observable (Fig. 7b). Such additional tests demonstrate again that the *SwarmCG* optimized parameters perform comparably to the Martini 3.0 ones even when transferred to model lipids not included in the training set.

## IV. CONCLUSIONS

We introduce an automated multi-objective FF optimization strategy and demonstrate its efficiency for developing optimized CG lipid FFs of variable resolution. The approach is general and can be used for optimizing explicit- or implicit-solvent FFs, with a variable resolution in the CG representation of the lipids. Newly devised OT-based metrics allow to quantify differences in the spatial organization of particles in molecular systems described at different resolutions. The parallel optimization of multiple lipid systems maximizes the transferability of the CG FF, and mitigates potential inaccuracies in the structure-based information (suboptimal AA FFs or limited MD sampling).

An intrinsic drawback of multi-parametric FF optimization is that multiple sets of parameters may produce the same results (*i.e.*, a high-dimensional problem is here formulated as a low-dimensional one). The transferability constraint, here induced by the fact that the building blocks of the FF (*bonded* and *non-bonded* interactions between CG particle types) are conserved in different lipid systems and are optimized in parallel, efficiently guides the optimization and discards spurious solutions. As demonstrated, this allows to build a solid framework of shared and sampled interactions, eventually allowing to transfer the optimized CG parameters to model other lipids. While an optimized CG FF becomes more complete as the diversity of the training set increases (more interactions between different beads become better sampled in the context of different molecular systems), our demonstrations indicate that the multiple interactions present in such complex bilayer systems guarantee a satisfactory transferability even using a reduced number of lipids in the training set. For the lipids within the training set, *SwarmCG* generates optimized FF parameters that systematically reproduce target experimental observables. As a proof-of-concept, here we optimize CG FFs for PC lipids described at different resolutions. At the same time, the obtained parameters can be transferred to other types of lipids, suggesting that substantial improvements could be achieved also in high-resolution models and even using limited data with this procedure. Increasing the diversity of the training set, including also other types of lipids, should allow to obtain even more general and accurate CG lipid FFs.

This automated multi-objective CG FF optimization strategy is general, and essentially requires that: (i) reference AA MD trajectories can be obtained for part of the lipids used in the training set (*bottom-up* requirement); (ii) reliable experimental data are available (*top-down* requirement); and (iii) the CG MD simulations used for testing the FF parameters are computationally accessible and sufficiently informative of the quality of the FF being optimized. Lipids thus constitute an appropriate use case, but the process can also be extended to parametrize FFs also for other classes of molecules for which reference experimental/simulation data are available. Because the form of the loss function used in this study is simple, it would be straightforward to input additional top-down observables to be evaluated in simulations, and to be quantified as percentages of deviation from target values (*i.e.* similar to modifying the loss function from Eq. 11 to Eq. 1). Moreover, this approach makes an efficient use of HPC resources and scales efficiently. In fact, while enlarging the training set may seem to increase the computational burden, this makes the transferability constraints more informative and the convergence of the optimization faster. The cost of the optimization process is fully compensated by the benefit, considered that optimized and transferable force fields can be obtained in a few days of calculation. Given the revolutions that transferable FFs brought to the molecular modeling community, we envisage that multi-objective optimization approaches, such as that presented here based on *SwarmCG*, will have great impact in the evolution towards next generation FFs.

## SUPPLEMENTARY MATERIAL

Supplementary material includes details on the functional form of the CG FFs for which the parameters are optimized, the various-resolution molecular models used in these experiments, their topologies, their optimized FF parameters obtained with *SwarmCG*, as well as the implementation for usage with HPC resources. Additional details are also provided concerning the submolecular features observed in the CG models obtained at the end of different FF optimization experiments, underlining the relevance of the OT-B and OT-NB metrics.

## ACKNOWLEDGEMENTS

G.M.P. acknowledges the funding received by the European Research Council (ERC) under the European Union's Horizon 2020 research and innovation program (grant agreement no. 818776 - DYNAPOL), by the Swiss National Science Foundation (SNSF grant: IZLIZ2_183336), and by H2020 under the FET Open RIA program (grant agreement no. 964386 - MIMICKEY). The authors also acknowledge the computational resources provided by the Swiss National Supercomputing Center (CSCS).

## AUTHOR DECLARATIONS

### Conflicts of Interest

The authors have no conflicts of interest to disclose.

### Author Contributions

C.E.M. and R.C. devised the OT-based metrics, the algorithm and built the molecular models. R.C., M.P., C.C. and G.D. performed the experiments. C.E.M. implemented the algorithm. C.E.M. and G.M.P. wrote the paper and designed the research. G.M.P. supervised the work. All authors agreed on the final form of the paper.

## DATA AVAILABILITY

The code used in this study for the optimization of the CG lipid models is available at: https://github.com/GMPavanLab/SwarmCGM, together with the models obtained at convergence for each experiment, as well as the configuration files used in this study, and all material necessary for running the software and for reproducibility testing.

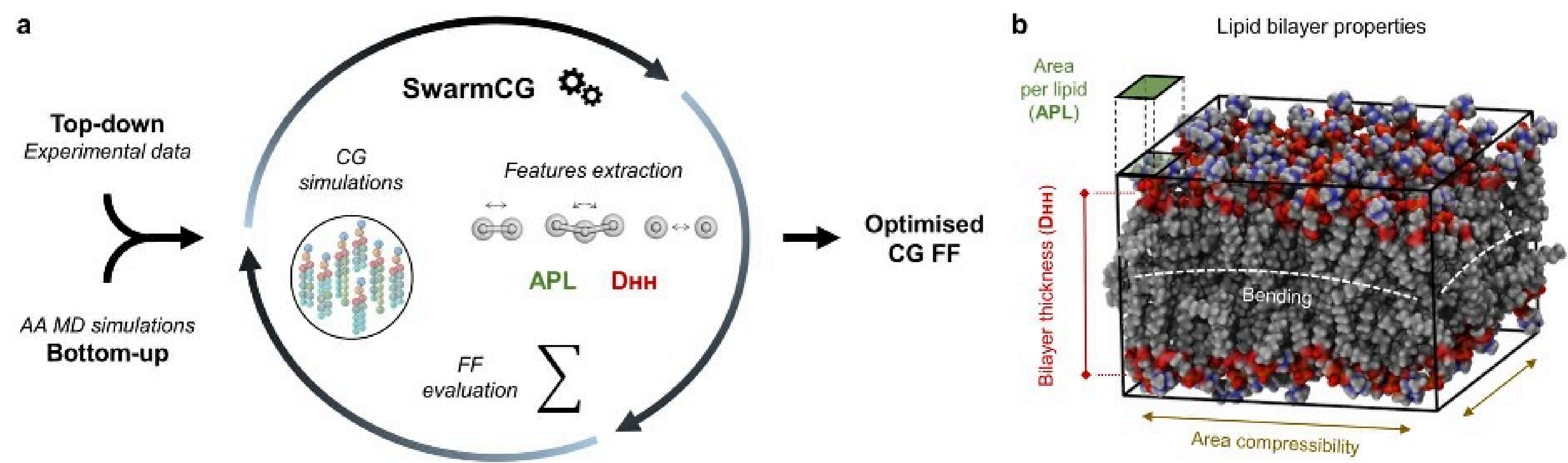

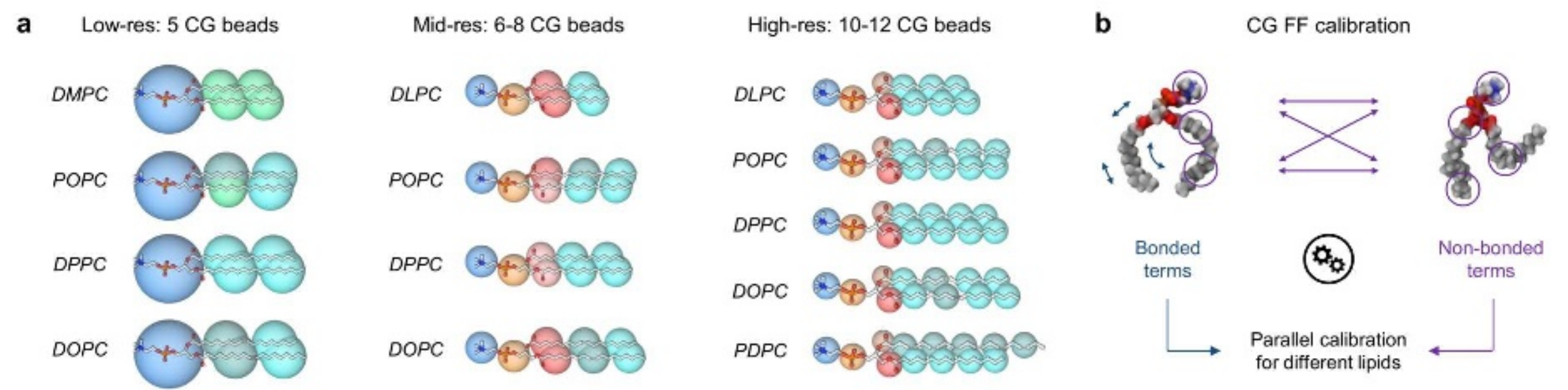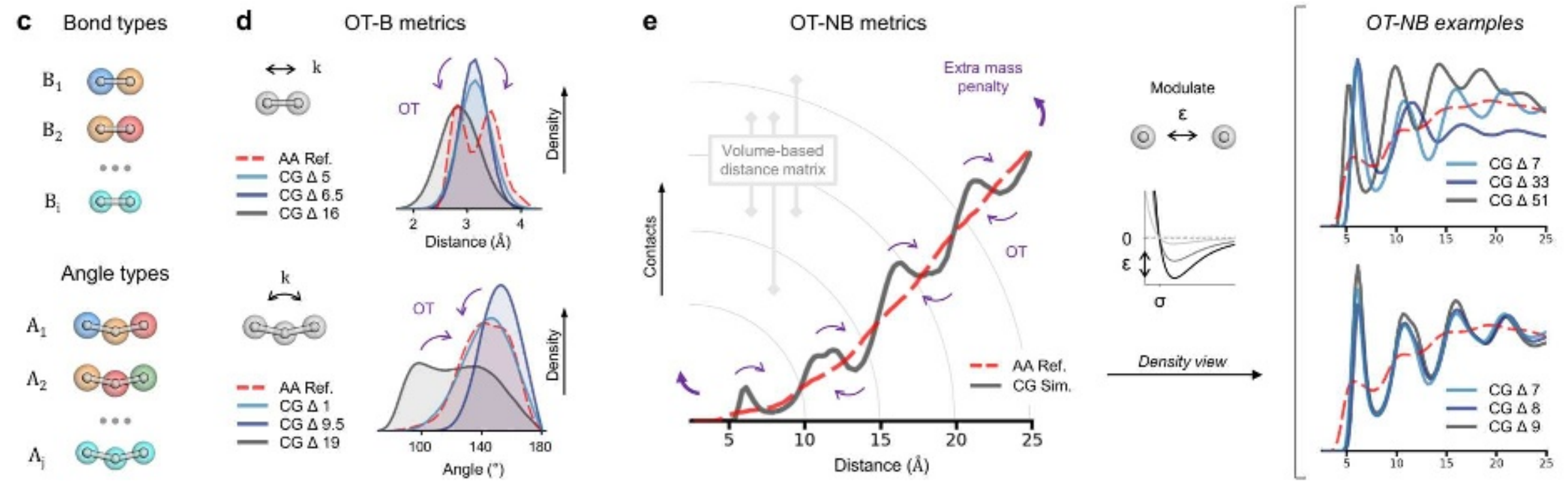

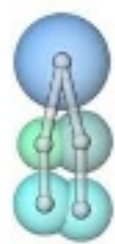 **a** 5 CG beads
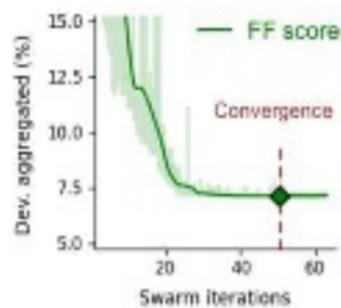
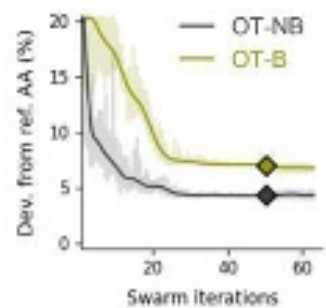
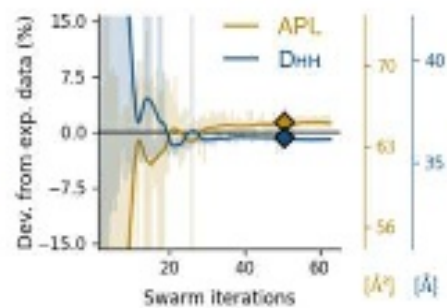
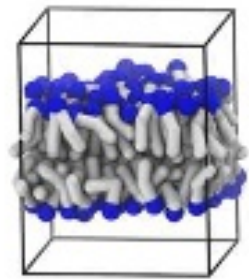

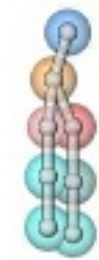 **b** 8 CG beads
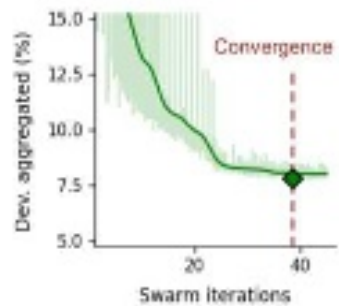
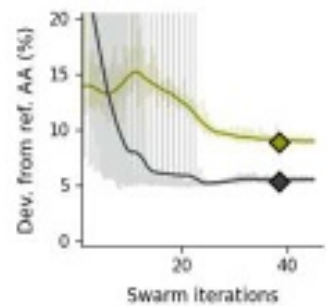
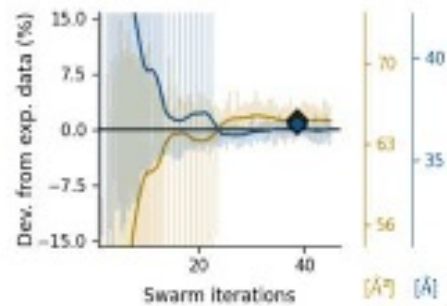
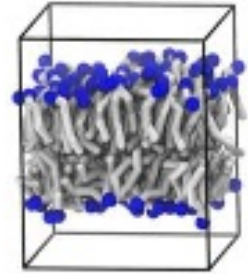

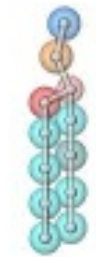 **c** 12 CG beads
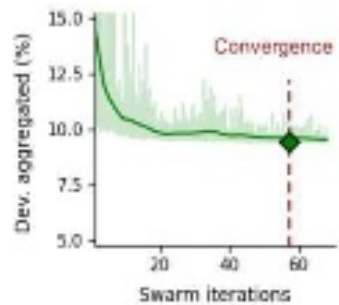
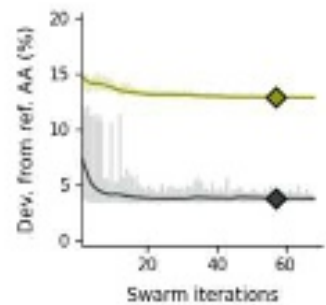
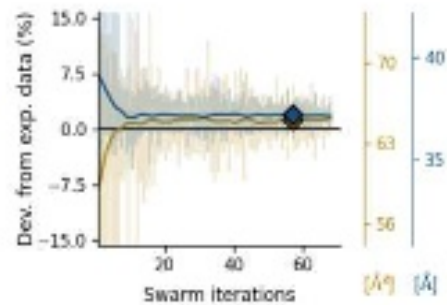
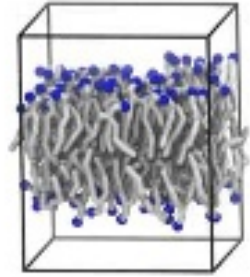

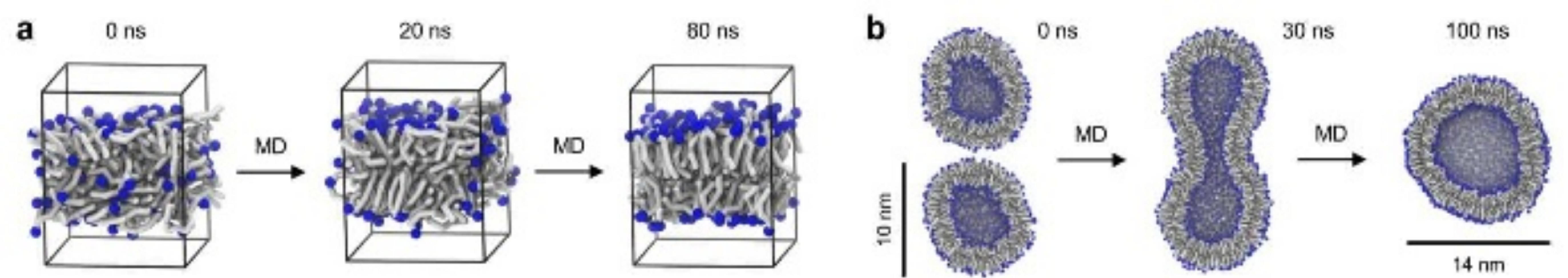

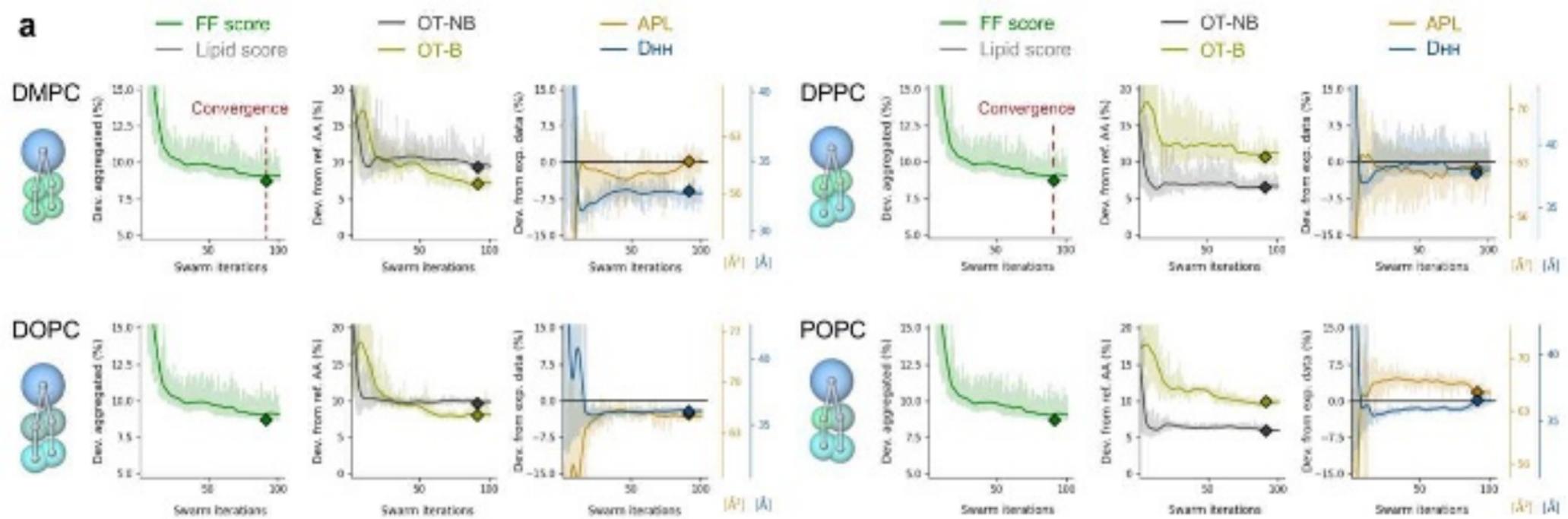
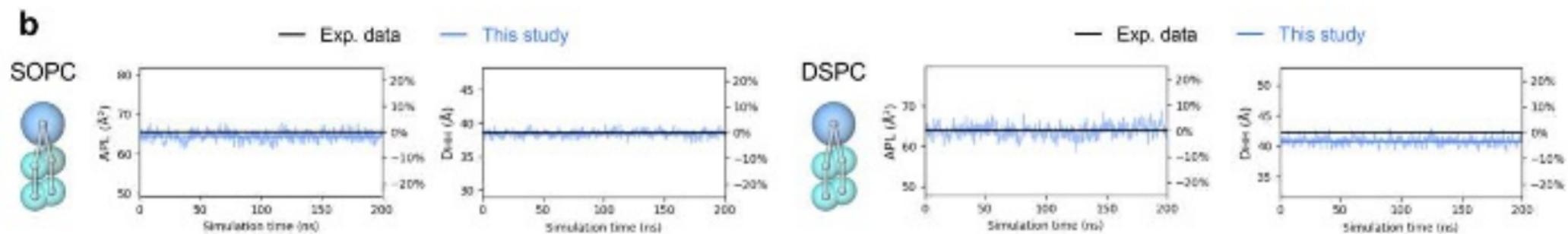

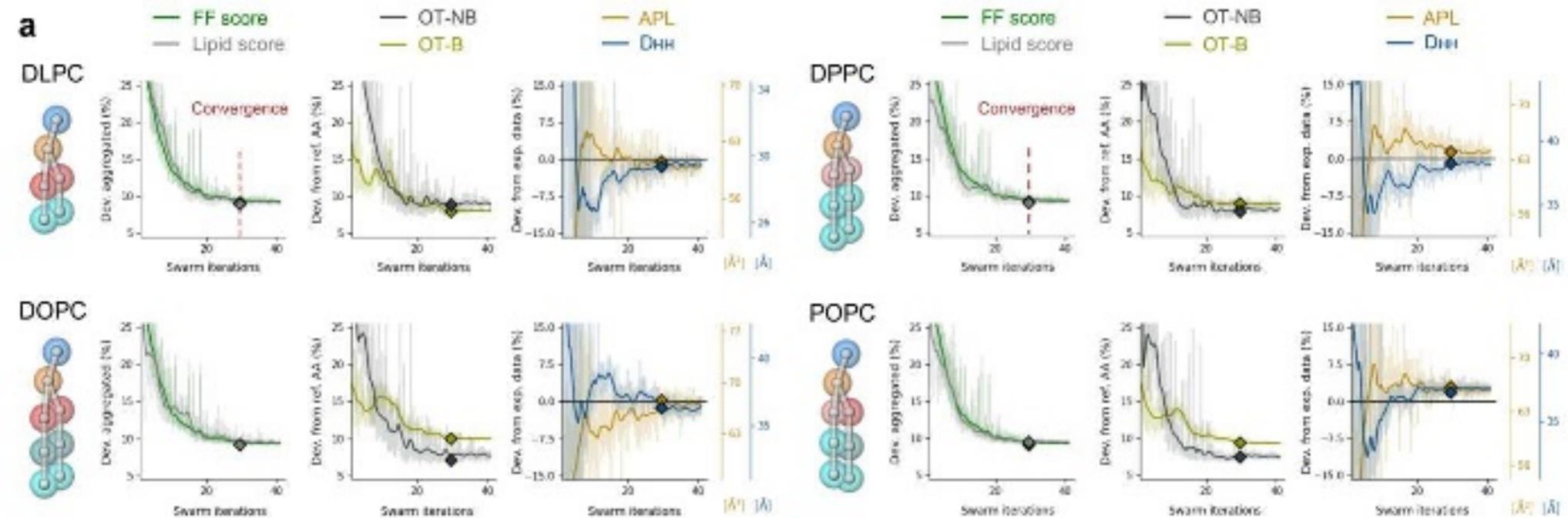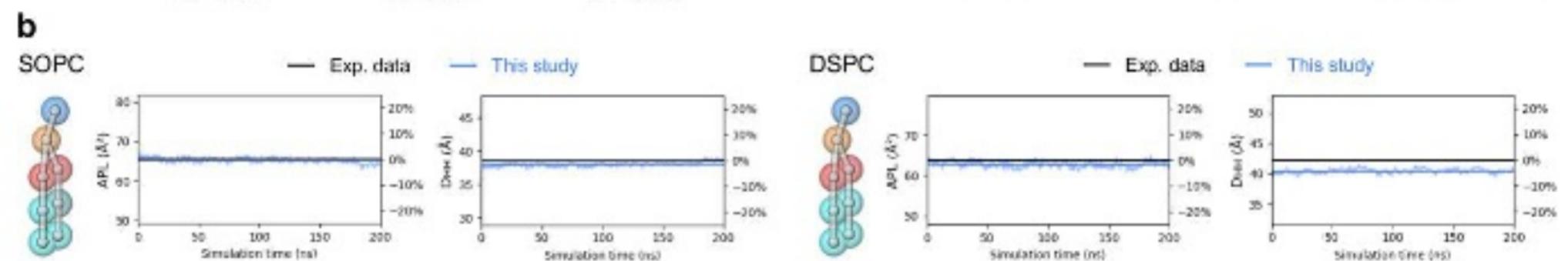

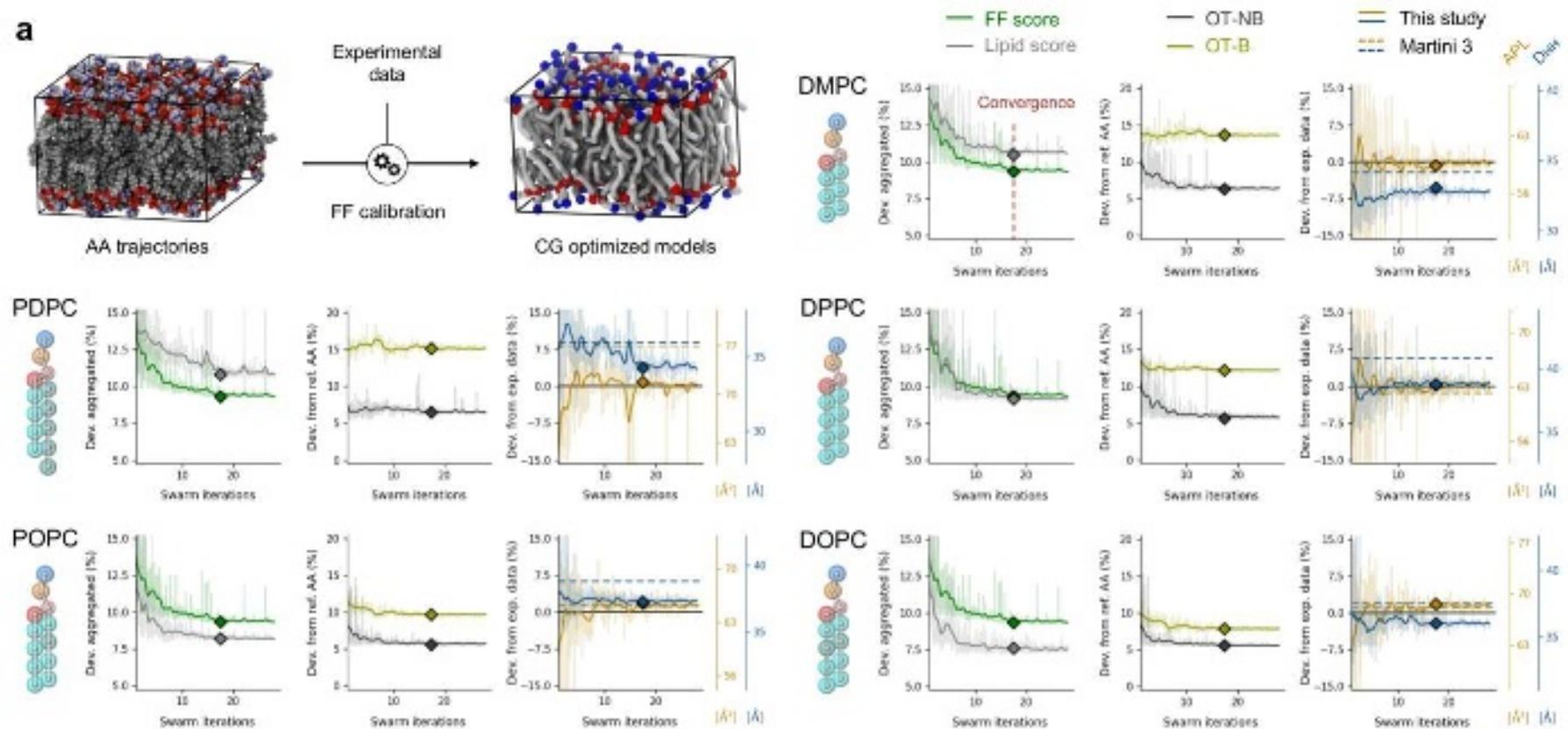
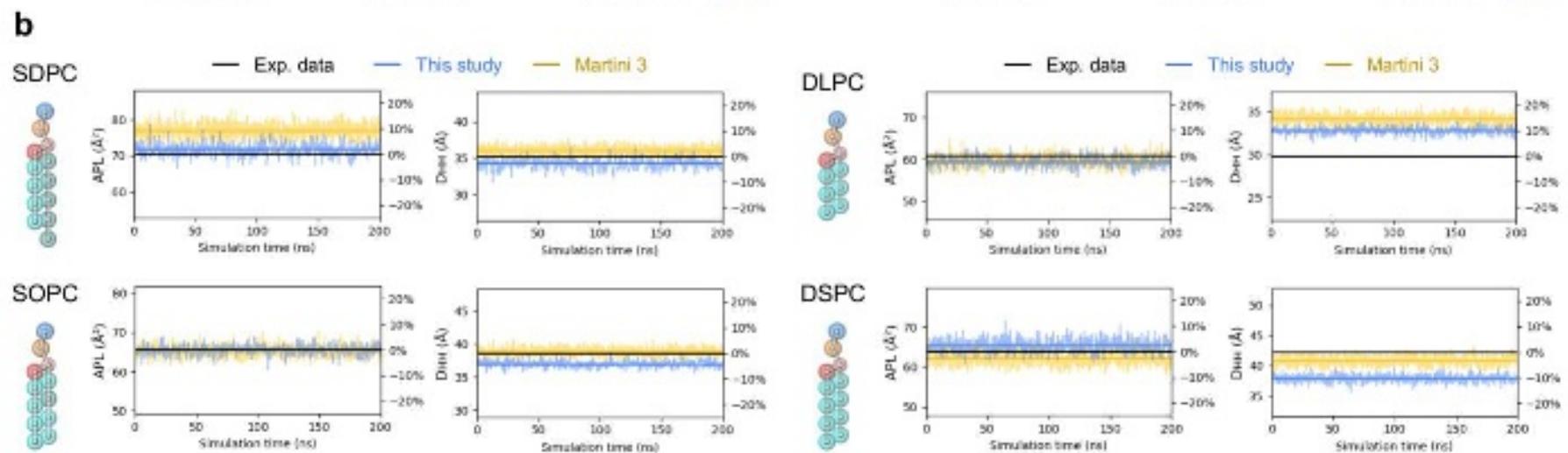